\documentclass{elsart}
                                                                                
\usepackage{graphicx}
                                                                                
\usepackage{amssymb}

\def\menorsim{\smash{\mathop{<}\limits_{\raise3pt\hbox{$\sim$}}}}
\def\maiorsim{\smash{\mathop{>}\limits_{\raise3pt\hbox{$\sim$}}}}

\begin{document}

\begin{frontmatter}

                                                                                

\title{Percolation and high energy cosmic rays above $10^{17}$~eV}


\author[sant]{J. Alvarez-Mu\~niz}
\author[ist]{P. Brogueira}
\author[lip]{R. Concei\c{c}\~ao}
\author[centra,ist]{J. Dias de Deus}
\author[lip,ist]{M.C. Esp\'{\i}rito Santo}
\author[lip,ist]{M. Pimenta}

\address[sant]{IGFAE and Dep. Fisica Particulas, Univ. Santiago de Compostela, 
15782 Santiago de Compostela, Spain}
\address[ist]{Departamento de F\'{\i}sica, IST, Av. Rovisco Pais, 1049-001 Lisboa, Portugal}
\address[lip]{LIP, Av. Elias Garcia, 14-1º, 1000-149 Lisboa, Portugal}
\address[centra]{CENTRA, Av. Rovisco Pais, 1049-001 Lisboa, Portugal}

\begin{abstract}
In this work we argue that, in the interpretation of the energy dependence of 
the depth of the shower maximum and of the muon content in high energy cosmic
ray showers ($E \maiorsim 10^{17}$~eV), other variables besides 
the composition may play an important role,
in particular those characterising the first (high energy) hadronic collisions. 
The role of the inelasticity, of the nature of the leading 
particle, and of the particle multiplicity are discussed. 
A consistent interpretation of existing data within a string percolation 
model implemented in a hybrid, one dimensional simulation method is given.

\end{abstract}

\end{frontmatter}

\section{Introduction}
\label{sec:introd}

The composition of cosmic rays at high energies is
a matter of controversy. Direct measurements are possible only up to 
$E \sim 10^{15}$~eV~\cite{shibata}. Above these energies, 
attempts to infer the effective mass number $A$ of the primary particle   
are based on the measured shower variables and are rather indirect~\cite{watson}.
Furthermore, data are relatively scarce at such high energies.

In fluorescence experiments, the longitudinal shower profile is measured and 
the atmospheric depth of the shower maximum $X_{max}$ is usually interpreted as 
an indication of 
the nature of the primary particle~\cite{watson&nagano}: 
heavy nuclei initiated showers are expected 
to develop earlier, due to the larger interaction cross-section.

In ground array experiments, the muon component is widely studied as 
a possible handle on composition~\cite{watson&nagano}. 
The muon content, or more specifically the muon density at a given 
distance from the shower core, $\rho_\mu$, is expected to increase by 
roughly 50\% from proton to Fe initiated showers, 
for the same primary energy~\cite{knapp}. 
Incidentally, the muon 
component of the shower is a powerful tool for the validation of hadronic
interaction models at such high energies.

In this work we argue that, in the interpretation of the energy dependence of 
$X_{max}$ and $\rho_\mu$, other variables besides $A$ may play an important role,
in particular those characterising the first (high energy) hadronic collisions. 
The role of the inelasticity $K$, one minus the fraction of momentum 
carried by the fastest (leading) particle, of the nature of the leading 
particle, and of the charged particle multiplicity are discussed. 
A consistent interpretation of data, within a string percolation model
implemented in a hybrid one dimensional simulation method, is given.

This paper is organised as follows.
In section~\ref{sec:exper} the experimental status is briefly reviewed.
In section~\ref{sec:model}, the possible role of the main variables
characterising the first hadronic collisions is analised
and the models and tools relevant for the discussion are presented.
In section~\ref{sec:results},
results on the energy dependence of $X_{max}$ and $\rho_\mu$
are derived
in the light of string percolation. In section~\ref{sec:summary}
some conclusions are drawn.

\section{Experimental status}
\label{sec:exper}

Experimental data on the energy dependence of the depth of the shower maximum 
$X_{max}$ and of the muon density 600 m away from the shower core $\rho_\mu(600)$
above $10^{17}$~eV, and their comparison with simulations, are summarised
in this section.

Fly's Eye/HiRes results in this energy range show an increase in the $X_{max}$
vs. $E$ slope, which has been interpreted as a change in composition,
going from more Fe-like to more proton-like showers~\cite{hires}. 
It has been argued~\cite{jonesETC,ourpaper} that other models 
predicting a development of the shower deeper in the atmosphere
(an increase in $X_{max}$) could also explain this effect.

Concerning the muon component of the shower, experimental data are
relatively scarce at high energies. Extensive air showers are 
largely electromagnetic dominated, and the separate detection of muons 
in ground array experiments requires the installation of specific shielded 
detector units. This has been done in KASCADE~\cite{kascade} 
in the energy range from $10^{14}$~eV to almost $10^{17}$~eV
and in AGASA above $10^{17}$~eV~\cite{agasa}. 
Hybrid data from HiRes-MIA has also been presented~\cite{mia}.

At AGASA, the lateral distribution function of muons above 0.5~GeV was measured
and combined with the Akeno 1~Km$^2$ array data (threshold 1 GeV)~\cite{agasa}.
The evolution of $\rho_\mu(600)$ with the particle density $S_0(600)$
(essentially proportional to the primary energy) was studied
and compared with CORSIKA~\cite{corsika} simulations 
using different hadronic models in~\cite{knapp}.
It was observed that the slope of $\rho_\mu(600)$ vs. $E$ in
data is flatter than in simulation with any hadronic model and primary 
composition.
Taking $\rho_\mu \propto E^\beta$, data gives 
$\beta=0.84 \pm 0.02$~\cite{knapp,agasa}, while 
from simulation  $\beta \sim 0.9$ ($\beta=0.92$ (0.89) 
for protons with
QGSJET (SIBYLL), while for Fe $\beta=0.88$ (0.87)~\cite{knapp}).
It has been argued~\cite{dawson} 
that these results can be interpreted as a change in composition
from heavy (at around $10^{17.5}$~eV) to light (at around $10^{19}$~eV),
agreeing with the Fly's Eye indication. It has been pointed out
in~\cite{knapp} that this interpretation should be studied in a wider 
energy range, as it seems to lead to a composition heavier than Fe at lower
energies.
The HiRes-Mia experiment consisted of the prototype high resolution 
Fly's Eye detector and the muon array MIA. Results
are presented in~\cite{mia}. The muon data, yielding
$\beta=0.73$, 
seem to indicate a composition heavier than Fe at $10^{17}$~eV.

\section{Construction of a simple model}
\label{sec:model}

\subsection{The composition interpretation}
\label{sec:comp}

In order to make the argument simple, we shall 
use the original Heitler idea~\cite{heitler}: 
the location of the shower maximum is, on the average, related to 
$\log E, \bar {X}_{max} \sim \log E$, and the number of charged pions or muons 
is proportional to some power of 
$E$, $N_{\pi^+ \pi^-} \sim N_{\mu} \sim E^{\beta}, 0 < \beta \menorsim 1$. 
In the case of a nucleus with $A$ nucleons colliding in a hadronic collision we shall write,
\begin{equation}
\bar X_{max} \simeq \bar X_1 + \bar X_0 \log (E/A) ,
\label{eq:Xmax-A}
\end{equation}
where $\bar X_1$ is the average depth of the first collision and $\bar X_0$ 
is the radiation length, and
\begin{equation}
\bar N_{\mu} \simeq A (E/A)^{\beta} ,
\label{eq:Nmu-A}
\end{equation}
or
\begin{equation}
\log N_{\mu} \simeq (1-\beta) \log A +\beta \log E \ . 
\end{equation}

We further have, for the elongation rate,
\begin{equation}
{d\bar X_{max} \over d\log E} = \bar X_0 \left[ 1- {d\log A \over d\log E}\right] ,
\label{eq:dXmax-A}
\end{equation}
and for the $\log N_{\mu}$ dependence on $E$,

\begin{equation}
{d\log N_{\mu} \over d\log E}=(1-\beta) {d\log A \over d\log E}+\beta \ .
\label{eq:dNmu-A}
\end{equation}

\noindent As experimentally~\cite{dova,hires,mia}, above $10^{17}$ eV $d\bar X_{max} /d\log E$ is larger 
and $d\log N_{\mu}/$ $d\log E$ is slightly smaller, in comparison with lower energies, 
the conclusion is:
\begin{equation}
\frac {d\log A} {d\log E} < 0 \ , 
\label{eq:dA}
\end{equation}
i.e., for $E \maiorsim 10^{17}$ eV the average mass number $A$ 
should, in this interpretation, decrease with energy.

\subsection{The first hadronic collisions interpretation}
\label{sec:K&P}

\begin{figure}[htb]
\begin{center}
\includegraphics[width=0.3\linewidth]{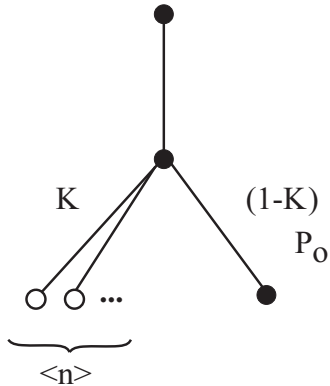}      
\end{center}
\caption{Schematic representation of the first hadronic collisions. A large fraction of the 
energy $(1-K)$ is taken by the leading particle. Most particles (essentially pions,
$<n>$ on average) come from the remaining energy, $K$.
The probability $P_0$ characterises the type of leading particle.}
\label{fig:Kscheme}
\end{figure}

In this interpretation a key role is given to the 
main variables characterising the first hadronic collisions, which are schematically
represented in Fig.~\ref{fig:Kscheme}.
The inelasticity $K$ is the fraction of energy distributed among 
the produced particles (mainly pions). A fraction of this energy is expected
to go into neutral pions, which promptly decay to photons.
$<n>$ is the average (non-leading) multiplicity at the collision energy. 
The quantity $(1-K)$ represents the fraction of energy concentrated in 
the leading particle, in general assumed to be a proton. $P_0$ is the probability
of having a leading $\pi^0$.

As pointed out in~\cite{jonesETC,gaisser2}, the effect of changing $A$ is equivalent to the effect 
of changing the average inelasticity $K$. 
In this spirit, 
the fastest particle, carrying an energy $(1-K)E$, will originate the shower branches that
go deeper in the atmosphere, and 
we write

\begin{equation}
\bar X_{max} \simeq \bar X_1 + \bar X_0 \log [(1-K){E\over E_0}] \ ,
\label{eq:Xmax}
\end{equation}
instead of~(\ref{eq:Xmax-A}), where $\bar X_1$ is the average depth of the first collision, $\bar X_0$ 
is the radiation length and $E_0$ a low energy threshold. 

Regarding the muon content of the shower, a possible assumption is to say that while the
energy flows in the $(1-K)$ direction (see Fig.~\ref{fig:Kscheme}) the number of particles
(of muons) flows in the $K$ direction. We thus have
\begin{equation}
N_\mu \sim N_\pi^\pm \propto KE \ .
\label{eq:Nmu}
\end{equation}
instead of~(\ref{eq:Nmu-A}).
The validity of this assumption will be discussed later.
We further have,
\begin{equation}
{d \bar X_{max} \over d\log E}= \bar X_0 \left[ {d\log (1-K) \over d\log E} +1\right] \ ,
\label{eq:dXmax}
\end{equation}
instead of~(\ref{eq:dXmax-A}), and
\begin{equation}
{d\log N_{\mu} \over d\log E}= {d\log K \over d\log E} + 1 \ ,
\label{eq:dNmu}
\end{equation}
instead of~(\ref{eq:dNmu-A}). The condition~(\ref{eq:dA}) becomes now 
\begin{equation}
\frac {d\log K} {d\log E} < 0 \ ,
\label{eq:dK}
\end{equation}
i.e., the inelasticity $K$ has to decrease with the energy.
This scenario has been explored in~\cite{ourpaper}, where a string percolation model
predicting the required behaviour of the inelasticity at high energies was discussed.
As it will be seen below, string percolation affects the energy dependence
of $K$, $P_0$ and $<n>$. 

\subsection{Matthews-Heitler toy model}
\label{sec:matthews}

In~\cite{matthews} a very interesting semi-empirical model for the development 
of the hadronic component of
air showers in analogy with Heitler's splitting approximation of electromagnetic 
cascades~\cite{heitler} 
was presented.
Not considering the elasticity, the model gives constant and reasonable
values for $dX_{max}/d\log E$ and  $d \log N_\mu /d\log E$. 
Taking into account that in hadronic interactions a significant fraction of the total
energy may be carried away by a single ``leading'' particle (i.e., $1-K>0$), the model predicts
for $N_\mu$ a slope decreasing with $K$ - exactly the opposite of~(\ref{eq:Nmu}).
It is interesting to discuss this discrepancy. 

In~\cite{matthews} the essential point is that the production of $\pi^0$'s, which decay
into photons before participating in the cascade, strongly decreases the muon 
production - about $1/3$ of the energy $KE$ is lost for muon production, as it goes 
into the electromagnetic component of the shower.
On the other hand, the leading particle, assumed in general to be a proton, goes on
producing particles, thus increasing the number of muons. 
However, if  $\pi^0$'s are themselves leading particles - $P_0>0$, 
as it happens in percolation models - the argument is no longer valid. 
In fact, when estimating muon production both the inelasticity $K$
and the probability $P_0$ of having a $\pi^0$ as a leading particle are
relevant. This will be seen next.

\subsection{Generalised Matthews-Heitler model}
\label{sec:toy}

The model of~\cite{matthews} 
was implemented and generalised. This is not a substitute for a detailed simulation, 
but useful to illustrate the physics involved.
Following~\cite{matthews},
hadrons interact after traversing one layer of fixed thickness (related to the
interaction length, $\lambda_I \sim 120$~g cm$^2$~\cite{matthews,gaisser}) 
producing $N_{ch}$ charged pions and  $N_{ch}/2$ neutral pions. 
A $\pi^0$ immediately decays into photons, meaning that $1/3$ of the
energy is lost to the electromagnetic shower component. Charged pions continue through
another layer and interact, until a critical energy is reached.
They are then assumed to decay, yielding muons, and thus $N_\mu = N_{\pi^\pm}$.
The version of the model which takes into account the inelasticity was implemented. 
In each branching, the energy was equally
shared amongst all (non-leading) particles. $N_{ch}=10$ was used.

\begin{figure}[hbtp]
\begin{center}
\includegraphics[width=0.8\linewidth]{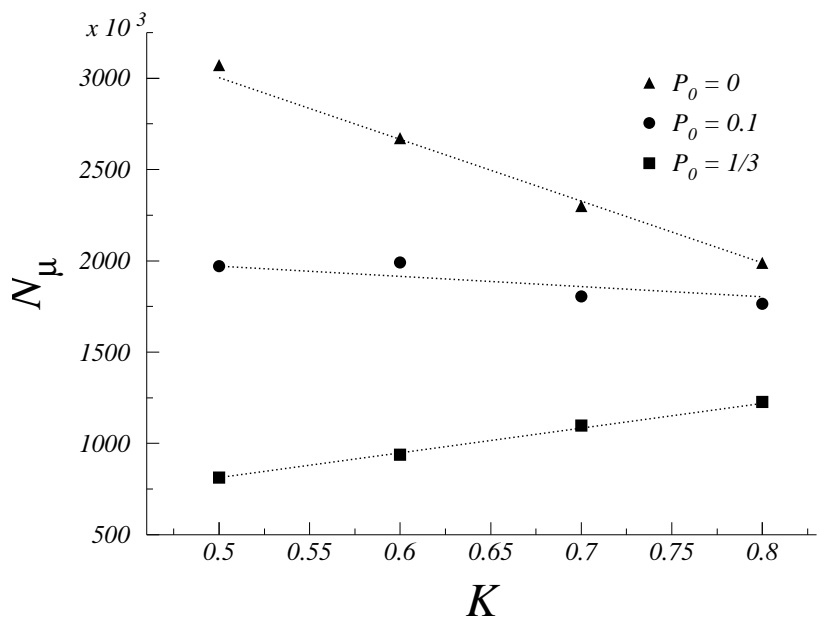}      
\end{center}
\caption{
The toy model prediction for the the number of muons as a function 
of the inelasticity is shown, for different values of the probability 
$P_0$ of having a leading $\pi^0$ and $E=10^{17}$~eV. 
The lines are only to guide the eye.
}
\label{fig:toy}
\end{figure}

We modified this model in two ways.
Firstly, a decay probability for charged pions in each interaction step was explicitly
included, replacing the large value of the critical energy (20 GeV) 
used in the original version. This gives a smoother and more realistic transition
from charged pions to muons. A critical energy of 1 GeV was then used. 
Further, the model was generalised to include the probability $P_0$ of having a leading
$\pi^0$ in a given collision, with an energy (1-K)E lost into the electromagnetic
branch in that collision. 
The obtained number of muons 
as a function of the inelasticity $K$ (for $E=10^{18}$~eV) 
is shown in Fig.~\ref{fig:toy}, for different values of $P_0$. 

As observed, in this toy model the $N_\mu$ slope depends
critically on the nature of the leading particle:
changing the probability $P_0$ of a leading neutral pion from 0 to 1/3 
inverts this slope. If the inelasticity is relatively low, even a moderate 
fraction of leading neutral pions may cause a decrease of the number
of produced muons. Thus, one easily moves from the behaviour predicted
in the Matthews-Heitler model (section~\ref{sec:matthews}) to the
first hadronic collisions model (section~\ref{sec:K&P}, equation~(\ref{eq:Nmu})).
It should however be noted that
in this simplified approach we do not take into account the fact that
the energy spectrum of the leading particle will depend on its nature:
leading neutral pions may be softer than leading protons.
More detailed simulations will be presented below.

\subsection{Hybrid shower simulations}
\label{sec:hybrid}

In order to validate the argument and derive predictions for the behaviour 
of $X_{max}$ and $\rho_\mu$ as a function of the energy, we need a shower
simulation tool which is detailed enough to produce reliable results, and
fast and flexible enough to allow the $ad-hoc$ introduction of the predictions
of the percolation model that will be used.

The hybrid, one dimensional simulation method described in~\cite{jaime}
was used.  It is a fast, one dimensional calculation, which provides predictions for the 
total number of charged particles and muons above several energy thresholds 
along the shower axis, as well as for the
fluctuations of the electromagnetic and hadronic components of the shower.
The method is based on precalculated showers (sub-threshold particles are treated with 
a library of profiles based on pre-simulated pion-initiated showers)
and a bootstrap procedure to extend the shower library to high energy.
The SIBYLL hadron interactions model was used.

\begin{figure}[hbtp]
\begin{center}
\includegraphics[width=0.8\linewidth]{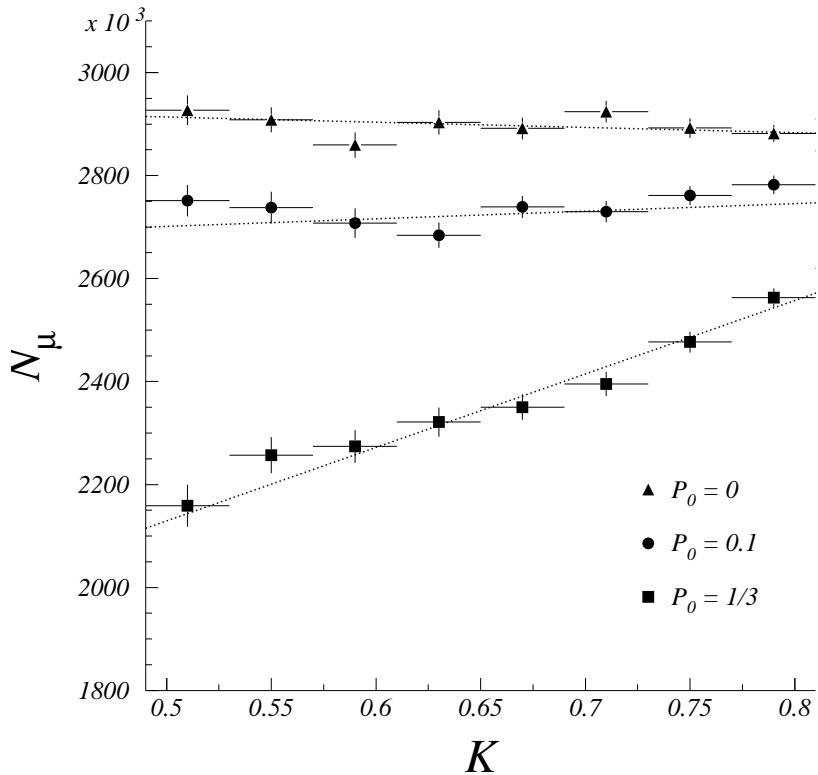}      
\end{center}
\caption{
The hybrid simulation prediction for the the number of muons 
($E>1$~GeV) as a function 
of the inelasticity is shown, for different values of the probability 
$P_0$ of having a leading $\pi^0$ and $E=10^{18}$~eV.
The lines are only to guide the eye.
}
\label{fig:hybrid}
\end{figure}

In Fig.~\ref{fig:hybrid}, the number of muons ($E>1$~GeV) as a function of $K$ is shown,
for  $E=10^{18}$~eV, and for different values of the fraction of
leading $\pi^0$'s, $P_0$. 
This confirms the indication obtained above with the toy model:
The slope of $N_\mu$ vs. $K$ depends strongly on $P_0$.

Studies performed with both QGSJET and SIBYLL show that the 
prediction of CORSIKA for $N_\mu$ vs. $K$ is
essentially flat.
In QGSJET, the fraction of leading $\pi^0$'s
in the first hadronic collision goes roughly from $10\%$ 
to $15\%$ as the primary energy goes from $10^{15}$~eV 
to $10^{20}$~eV, but their spectrum is relatively soft.
The variation of $P_0$ in SIBYLL is very small (of the order of
1\% for the same energy range).

The combined effect of $K$ and $P_0$ has thus an effect on muon production,
and this is particularly true if the inelasticity has relatively low values.
We now need a consistent model with predictions for the values of these
variables.

\subsection{String percolation model}
\label{sec:percol}

Essentially, all existing high energy strong interaction models based on QCD, 
and QCD evolution, predict an increase with energy -- not a decrease -- of the 
inelasticity $K$~\cite{qcd}. The same is true for the hadronic generators SIBYLL~\cite{sibyll} 
and QGSJET~\cite{qgsjet}, used in cosmic ray cascade analysis. This happens because 
evolution in the energy implies transfer of energy from valence partons or strings, 
or from bare Pomeron diagrams, to sea partons or strings, or to multi-Pomeron contributions.

However, in models with percolation of partons or strings, one expects the inelasticity $K$ 
to decrease with energy above the percolation threshold~\cite{ourpaper}. 
In the framework of the Dual String Model~\cite{dualstring} -- but we believe the argument is more 
general -- what happens at low energy is the transfer of energy from the valence strings 
to sea strings (and $K$ increases), while at higher energy the strings start to overlap 
and a cumulative effect occurs: the length in rapidity of fused (percolated) strings 
is larger~\cite{percol1}. At some stage, close to percolation threshold, the percolating strings take 
over the valence strings, and from then on $K$ decreases with the energy. Percolation is, 
in fact, a mechanism for generating fast leading particles.

\begin{figure}[hbtp]
\begin{center}
\includegraphics[width=1.0\linewidth]{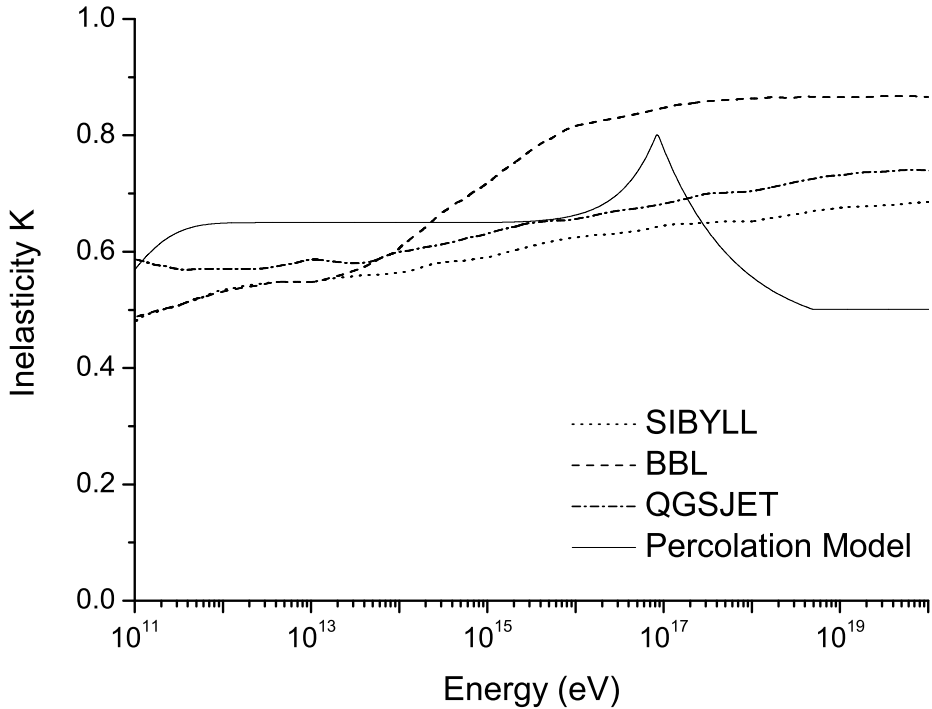}      
\end{center}
\caption{Energy dependence of the average inelasticity $K$. 
Conventional QCD model results are from~\cite{drescher}: 
dotted curve is SIBYLL simulation; dashed-dotted curve is QGSJET simulations; 
dashed is a model of~\cite{drescher}; 
full line is our percolation model.}
\label{fig:Kpercol}
\end{figure}

In Fig.~\ref{fig:Kpercol} we show the energy dependence of $K$ in the case of our string 
percolation model~\cite{ourpaper}, 
in comparison with $K$ determined from QCD inspired models, without percolation (see for 
instance~\cite{drescher} ). 
As discussed in~\cite{ourpaper}, this model was tuned taking into account multiplicity 
data from accelerator energies and places the percolation threshold
at $E \simeq 10^{17}$~eV. 
In the reference, predictions for the inelasticity and multiplicity behaviour were made  
and their effect on $X_{max}$ data was discussed.
Above the percolation threshold the inverted slope of $K$ vs. $E$
is clearly visible.
In the relevant energy range, $K$ varies roughly from 0.8 to 0.55.
Percolation will also affect the particle multiplicity,
and we follow the treatment in~\cite{ourpaper}, where we see that this effect 
can be introduced through the colour summation reduction factor $F( \eta )$, where $\eta$ is 
the transverse density.
 
Let us now turn to the evolution with energy of the probability $P_0$ of having a 
non-baryonic primary.
As energy increases, the number of sea strings increases. These are, however, low energy
strings, and, without percolation, the initial proton remains dominant.
When sea strings percolate, larger ``sea clusters'' are formed, and other particles
can be produced. It is a prediction of our percolation model that above the percolation
threshold, because sea strings are of the type quark-antiquark, 
the probability of having a leading $\pi^0$ will tend to $1/3$. 

\section{Results}
\label{sec:results}

We now have a string percolation model which predicts modifications on the variables
$K$, $P_0$ and $<n>$ characterising hadronic collisions above a percolation threshold. 
Inserting these predictions into the hybrid shower simulation tool described in
section~\ref{sec:hybrid},
results for the behaviour of $N_\mu$ and $X_{max}$ as a function
of the energy can now be derived and compared with the available 
data (see section~\ref{sec:exper}).

The percolation prediction for the inelasticity $K$ was introduced
in the hybrid simulations for hadronic collisions above $10^{16}$~eV. 
Above the percolation threshold ($E \sim 10^{17}$~eV),
the probability of having a leading $\pi^0$ of $P_0$  was set
to $1/3$ and the percolation multiplicity reduction factor $F(\eta)$ 
was introduced.

\begin{figure}[hbtp]
\begin{center}
\includegraphics[width=0.6\linewidth]{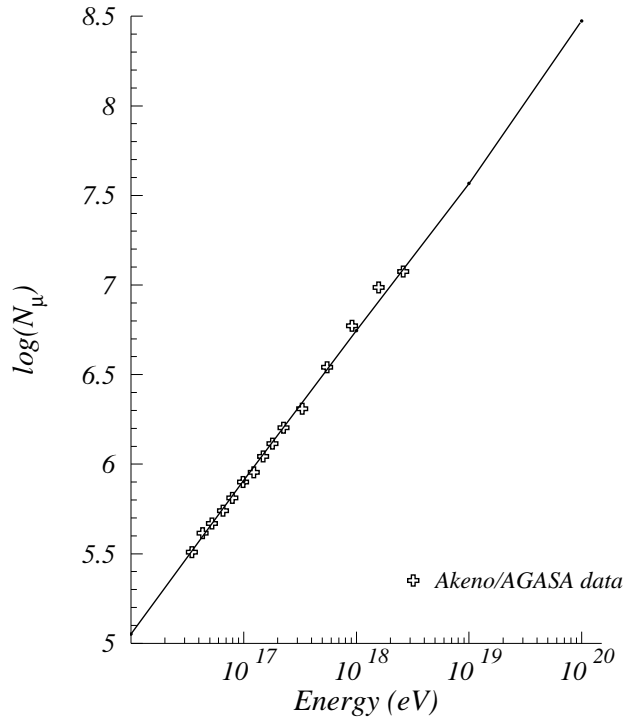}      
\end{center}
\caption{
The number of muons as a function of the primary energy.
The results of our model (full line) are shown and compared with Akeno/AGASA data (symbols).
}
\label{fig:results-mu}
\end{figure}

In Fig.~\ref{fig:results-mu} our results for $N_\mu$ vs. $E$ are shown and
compared with Akeno/AGASA data~\cite{agasa,agasa2}.
A value of $\beta = 0.83$ is obtained from simulation,
in very good agreement with the measured value of $\beta=0.84\pm0.02$.
In this study a primary proton was considered.
The result is however more general, as 
it has been shown in~\cite{knapp} that, for fixed composition, the value of 
$\beta$ is essentially the same for proton and for Fe.
The curve shown in the figure corresponds to shifting our proton result,
using equation~(\ref{eq:Nmu-A}),
to an intermediate and constant composition $A \sim 20$. 
The result of $\beta=0.73$ quoted by HiRes-Mia can hardly
be accommodated in this model, as it would imply values of 
$P_0$ going well beyond the expectations. 

It is interesting to note that at very high energies ($E \maiorsim 10^{19}$~eV)
the $\pi^0$'s starting
to participate in the shower (i.e. to interact) and hence contribute to the
production of muons. This causes an increase in the slope. A slight effect
is already seen in our simulations. 
Finally, as muon data is often presented as muon density $\rho_\mu(600)$ vs. $E$, 
it is worth noting that CORSIKA~\cite{corsika} was used to check that 
$\rho_\mu(600)$ behaves much like $N_\mu$ as a function of the energy (with $N_\mu$
slightly steeper, but differences of the order of 0.02).

\begin{figure}[hbtp]
\begin{center}
\includegraphics[width=0.6\linewidth]{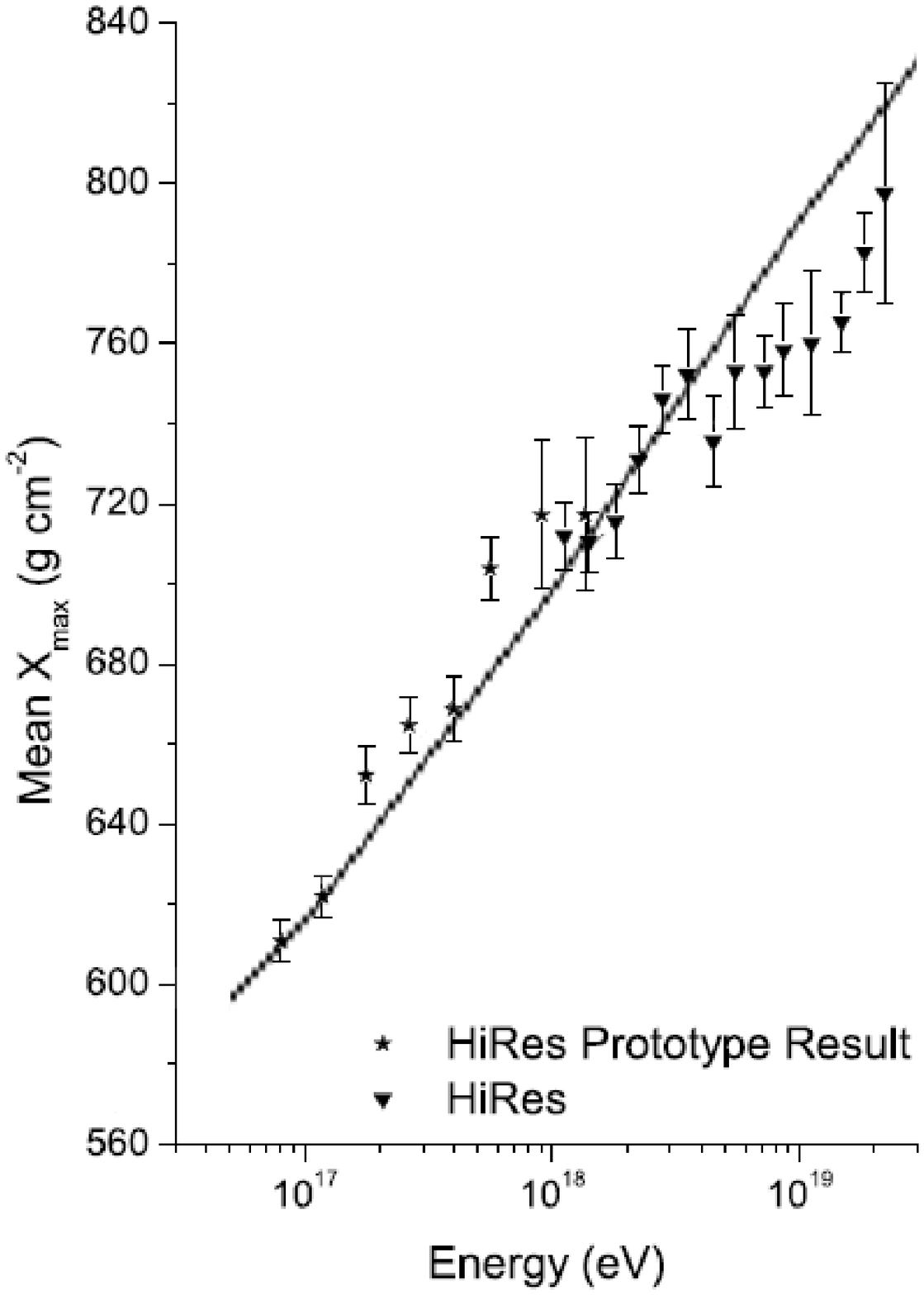}      
\end{center}
\caption{
The depth of the shower maximum as a function of the primary energy.
The results of our model (full line) are shown and compared with HiRes data (symbols).
}
\label{fig:results-xmax}
\end{figure}

In Fig.~\ref{fig:results-xmax} the results obtained for $X_{max}$ vs. $E$ are shown and
compared with existing data. A composition very similar to the one above was 
used. We see that the obtained curve is reasonably consistent with data and with
the result presented in~\cite{ourpaper} using a simple model similar to the one
described in section~\ref{sec:K&P}. 

We thus conclude that a percolation model with predictions
for the evolution with energy of the inelasticity $K$, the multiplicity $<n>$
and the fraction $P_0$ of leading neutral pions, in high energy
hadronic collisions, can contribute to explain consistently 
the behaviour observed in data,
both for $N_\mu$ vs. $E$ and $X_{max}$ vs. $E$.

\section{Summary and conclusions}
\label{sec:summary}

In this paper we have shown that, in the interpretation of the energy dependence of 
the depth of the shower maximum and of the muon content in high energy cosmic
ray showers ($E \maiorsim 10^{17}$~eV), other variables besides 
the composition may play an important role, in particular those
characterising the first (high energy) hadronic collisions.
The inelasticity $K$, the nature of the leading 
particle $P_0$, and the particle multiplicity $<n>$ were discussed. 

We developed a model which includes a hybrid one-dimensional
Monte Carlo simulation~\cite{jaime} and expectations from the percolation 
scenario~\cite{ourpaper}. 
For the first time the predictions of a percolation model for the different
variables were included systematically and used to derive predictions
on the behaviour of the depth of the shower maximum $X_{max}$ and on the
number of muons in the shower $N_\mu$ as a function of the energy.
We were able to explain the data in a consistent way.
The model reasonably describes the trends seen in data.

It shoud be noted however that the model (in particular the behaviour of the number
of strings as a function of the energy~\cite{ourpaper}) is tuned for high 
energy, the region where the percolation threshold arises ($E \maiorsim 10^{17}$~eV). 
The next step will be to create a full Monte Carlo simulation including percolation, 
adapted to the low and high energy regions.
This is particularly important for the case of $N_\mu$, due 
to the influence of $P_0$. A detailed Monte Carlo
calculation is thus required for a better understanding of data at all
energies. This will allow to further test the predictions
of percolation, namely the relative decrease of multiplicity, the non
monotonical behaviour of the inelasticity and the fast growth of $P_0$ with
energy, around $10^{17}$~eV. 

Finally, and concerning the leading $\pi^0$'s, it is important to realise that above
$10^{19}$~eV they become active in the cascade, thus generating more
hadrons. The $\beta$ parameter is expected to increase, and this is already
seen in our simulations. However, a proper
treatment of this problem requires, once again, a detailed Monte Carlo on 
percolation. More data at extremely high energies are also needed.

\section*{Acknowledgements}

We thank C. Pajares and S. Andringa for useful discussions.
J. A.-M. is supported by the ``Ram\'on y Cajal'' program and thanks LIP for 
the hospitality.  

 

\end{document}